\documentstyle[twocolumn,pre,aps]{revtex}
\begin{document}
\twocolumn[\hsize\textwidth\columnwidth\hsize\csname @twocolumnfalse\endcsname
\title{Hydrodynamics of Monolayer Domains \\ at the Air-Water Interface}

\author{David K. Lubensky\thanks{email: lubensky@fas.harvard.edu}$^{\natural}$}

\address{Department of Physics, Joseph Henry Laboratories,
Princeton University, Princeton, NJ 08544}
\address{Institut Charles Sadron, 6 rue Boussingault, 67083 Strasbourg Cedex,
France}

\author{Raymond E. Goldstein\thanks{email: gold@davinci.princeton.edu}}

\address{Department of Physics, Joseph Henry Laboratories,
Princeton University, Princeton, NJ 08544}

\maketitle
\begin{abstract}

Molecules at the air-water interface often form inhomogeneous
layers in which domains of different densities are separated by sharp
interfaces.  Complex interfacial pattern formation may occur through the
competition of short- and long-range forces acting within
the monolayer.  The overdamped hydrodynamics of such interfacial motion
is treated here in a general manner that accounts for 
dissipation both within the monolayer and in the subfluid. 
Previous results on the linear stability of interfaces are
recovered and extended, and a formulation applicable to the nonlinear regime is
developed.   A simplified dynamical law valid when dissipation in the monolayer
itself is negligible is also proposed.  Throughout the analysis, special
attention is paid to the dependence of the dynamical behavior on a 
characteristic length scale set by the ratio of the viscosities 
in the monolayer and in the subphase.

\end{abstract}

\vskip2pc]

\section{Introduction and Experimental Background}

Molecular layers of surfactants or polymers at the air-water interface are
often found in inhomogeneous states within which appear
domains of nearly uniform density~\cite{mohwald,mcc-anpc}.
In many ways, these states resemble conventional two-phase coexistence regions
with sharp interfaces between different homogeneous phases;
because of long-range electrostatic interactions within the monolayer, however,
domains of a given phase
can be stable rather than coarsening in time~\cite{Seul_Andelman}.
Experimentally observed domains typically have sizes of $10-100$ $\mu{\rm m}$.

A considerable body of
experimental~\cite{ss,sp,sknob,bmc,mlange,seul-dyn,stine-strat,lee-mcc,emann}
and theoretical~\cite{stone-mcc,stone-mcc2,gold-mono} work has focused on the
motion of the
domain boundaries. These investigations have had two complementary
motivations.  On the one hand, the boundary dynamics provide a means of probing
physical parameters, such as the line tension between phases, that are
otherwise difficult to measure~\cite{bmc,emann,gold-mono}.  On the other,
the electrostatic interactions are caused by
the molecules' permanent dipoles that are oriented
with respect to surface of the water; their mutual repulsion can result in
intricate fingering instabilities that have parallels in a variety  of other
pattern-forming systems \cite{Seul_Andelman}.  The laws of motion of the
interfaces between monolayer phases have been probed directly in experiments
monitoring the relaxation of domains to a circular ground state, starting
either
from an elongated ``bola'' shape~\cite{bmc,emann} or from smaller elliptical
deformations~\cite{seul-dyn,emann}.  The fastest-growing mode at the onset of a
fingering instability has also been
examined~\cite{ss,sp,lee-mcc}.

The complete interpretation of
such experiments requires an understanding of the hydrodynamics of thin layers
coupled to a subfluid.  Some progress has been made in the analysis of
particular cases.
Building on earlier work on the diffusion of a cylinder imbedded in a
membrane~\cite{hpw}, Stone and McConnell have solved the linearized boundary
dynamics about a circle~\cite{stone-mcc,stone-mcc2}.  Schwartz, Knobler, and
Bruinsma~\cite{skb} and Stone~\cite{stone}  have examined the flow of
monolayers
through channels.  The related problem of fingering in quasi-two-dimensional
domains of ferrofluids has also recently been
considered~\cite{cebers-m,cebers-z,gold-pra,gold-sci,gold-tse}, as has the
behavior of capillary
waves in the presence of coexisting monolayer phases~\cite{chou-nels}.
The more general treatment of boundary motion applicable to a domain
of any shape has, however, remained an open problem of considerable interest.

In this paper, we adapt boundary integral techniques from fluid
mechanics~\cite{pozr.bk} to the study of thin layers resting on a
subfluid.  Our approach is applicable to arbitrary geometries, and hence offers
several advantages over techniques based on eigenfunction expansions
that are useful only in particular situations with a high degree of symmetry.
It allows the comparatively straightforward calculation of linearized growth
rates about any number of stable shapes; it is
also a starting point for the detailed investigation of the boundary
dynamics in the nonlinear regime.  Further, by separating those aspects of the
problem that depend on energetics from those determined by the
hydrodynamic equations, we gain several physical insights.  Finally, our
formulation provides a new example of a
dynamical law governing curve motion in the plane.  It is thus of
interest to the broader study of pattern formation.

In Section II, we present a formulation of the boundary dynamics
valid for arbitrary viscosities and subfluid depths, and note some global
properties of the resulting dynamical law.  Section III
considers in greater detail the more analytically tractable limiting
cases of an infinitely deep or very thin subfluid.  In
Section~\ref{linstab-sect}, we examine the linear stability of straight and
circular boundaries.  Drawing on these results, Section~\ref{large-a-sect}
proposes a
simplified dynamics appropriate when dissipation occurs mostly in the subfluid.
This limiting form has the intriguing feature that it contains a local drag
term at the interface not unlike that introduced in the Rouse model of
polymer dynamics~\cite{doi-ed}.
Section~\ref{perimeter-sect} gives a brief, illustrative calculation in the
nonlinear regime, and Section~\ref{discussion-sect} discusses some limitations
and possible extensions of our work.

\section{General Formulation}
\subsection{The Model and a Boundary Integral Formulation}

We begin by ignoring the presence of domains and studying a
homogeneous two-dimensional layer coupled to a water subphase. The geometry of
the system is shown in Fig. \ref{figure1}.  Both fluids are
taken to have an infinite horizontal extent, while the subphase may have a
finite depth $d$.  The monolayer is assumed to be an incompressible,
Newtonian fluid with surface viscosity $\eta$
(dimensions mass/time) filling the plane $z=0$~\cite{end-notea}.  It
rests on a three-dimensional incompressible fluid of viscosity $\eta'$
(dimensions
mass/(length$\times$time)) that occupies the region $-d<z<0$.  All variables
referring to the subfluid are primed.  Since the Reynolds numbers involved in
the
slow relaxation of micron-scale domains seldom exceed $10^{-4}$, we are
justified in working in the overdamped limit. Neglecting all inertial terms in
the
Navier-Stokes equation, we then find that the system is governed by the two
coupled Stokes equations
\begin{equation}
\eta \nabla^2 {\bf u}- \nabla p + {\bf F}^S = {\bf 0} ~,
\label{modstokes}
\end{equation}
and
\begin{equation}
\eta' \nabla^2 {\bf u}' - \nabla p' = {\bf 0} ~, \label{substokes}
\end{equation}
along with the
incompressibility conditions $\bbox{{\nabla}}\cdot {\bf u} = 0$ and
$\bbox{\nabla}\cdot {\bf u}'= 0$.
The term
${\bf F}^S$ gives the body force that the subfluid exerts on the monolayer.
Note that the monolayer is treated as a truly two-dimensional fluid, so $p$
is a surface pressure with dimensions of force/length.
The two
fluids are also linked by no-slip boundary conditions; these simply require
that,
at $z=0$, ${\bf u}'_{\perp} = {\bf u}$ and ${\bf u}' \cdot \hat {\bf z} = 0$,
where the $\perp$ indicates the in-plane component of a three-dimensional
vector.  In
addition, we demand that all velocities vanish at infinity, and we impose
no-slip
boundary conditions on the bottom of the subfluid trough.  Our model is
thus essentially the same as that first introduced by Saffman to model flow
in fluid membranes~\cite{saffman};
its predictions have been shown to agree
well with experiment for monolayers flowing through a channel~\cite{skb}.
The form of the force ${\bf F}^S$ is determined by the
subfluid's Newtonian stress tensor evaluated at $z=0$.  The no-slip boundary
conditions on the velocity field ${\bf u}'$ imply that all terms containing
${\bf u}' \cdot \hat{\bf z}$ or its derivatives vanish, leaving the simple
expression
\begin{equation}
{\bf F}^S = -\eta'{\partial {\bf u}'_{\perp}\over {\partial
z}}\Biggr\vert_{z=0}~.  \label{fsdef}
\end{equation}

The presence of a domain boundary will modify the equations for flow in a
homogenous
layer.  We describe the interface as a closed curve ${\cal C}$ in the $x-y$
plane
(Fig. \ref{figure1}) and ask that the fluid velocity be continuous across this
boundary. In order to make the problem more tractable, we also assume that the
two monolayer phases separated by the interface possess the same viscosity;
this
assumption will be discussed further later in this section.  With the curve is
associated a parametrization ${\bf r}(\alpha)$
and an energy ${\cal E}[{\bf r}]$ that is a functional of ${\bf r}(\alpha)$.
In
the simplest case in which there exists only a line tension $\gamma$, for
example, ${\cal E}[{\bf r}] = \gamma \int\! ds$, where $ds=\sqrt{g} d\alpha$,
and $\sqrt{g} \equiv  \vert d{\bf r}/d\alpha\vert$ is the metric factor.

To understand the effect of an interfacial force, we next
introduce the Green's function $G_{ij}$ that gives the response of a
monolayer {\em coupled to a subfluid} to a point force exerted on the
monolayer.
A force $\bf g$ acting at the origin will induce a velocity field $\bf u^{\rm
g}$
in the monolayer that is related to the Green's function by
\begin{equation}
u_i^{\rm g}({\bf r}) = {1\over 4\pi\eta}G_{ij}({\bf r}) g_j~,
\end{equation}
where summation over repeated indices is implied.  One can similarly
introduce Green's functions for the pressure and the body force
exerted by the subfluid, defined by the relations
\begin{equation}
p^{\rm g}({\bf r}) = \frac{1}{4 \pi} P_j({\bf r}) g_j \; \; \; {\rm and}
\; \; \; F_i^{\rm S,g}({\bf r}) = \frac{1}{4 \pi} f_{ij}^{\rm S}({\bf r}) g_j
\; .
\end{equation}
Together, the three will satisfy the equations
\begin{mathletters}
\begin{equation}
\nabla^2 G_{ij} - \partial_i P_j + f^{\rm S}_{ij} = 
-4\pi\delta_{ij}\delta({\bf r})
\label{forcedeqn}
\end{equation}
\begin{equation}
\partial_i G_{ij}=0~.
\end{equation}
\end{mathletters}
The first of these equations is the analog for Green's functions of
equation~(\ref{modstokes}) governing flow in the monolayer, while the second
reproduces an incompressibility constraint.  One could of course also write
down the equations corresponding to the
subfluid Stokes equation~(\ref{substokes}) and to the expression~(\ref{fsdef})
for $\bf F^{\rm S}$, but they are not necessary for the further development of
the present paper.

For our purposes, the essential feature of an interface is that it exerts a
force on the surrounding fluid.  To find the velocity field in the presence of
an interface, one must thus sum the contributions from the forces it exerts at
each point.  In other words,
\begin{equation}
u_j({\bf r}) = -{1\over 4 \pi \eta} \int_{\cal C} ds \, G_{ij}\left({\bf
r}-{\bf r}(s)\right) \Delta f_i({\bf r}(s))~,
\label{boundint}
\end{equation}
where we have chosen the sign convention that ${\bf \Delta f}$ is the net force
per
unit length that the fluid exerts on the interface.
With equation~(\ref{boundint}), we have effectively reduced the problem of
solving the two coupled partial differential equations~(\ref{modstokes})
and~(\ref{substokes})
to that of finding the correct Green's function;
although equation~(\ref{boundint}) makes no explicit reference to the subfluid,
it is included implicitly because its presence determines the form of
$G_{ij}$.
Virtual work arguments give an expression for ${\bf \Delta f}$ in terms of
${\cal E}[{\bf r}]$; when a small element of the surface is moved slightly,
the stresses in the surrounding molecular layer must do work to change  ${\cal
E}$.
The balance is expressed by the relation
\begin{equation}
{\bf \Delta f} = {1\over \sqrt{g}} {\delta {\cal E}\over \delta {\bf r}}~,
\end{equation}
which is simply a generalization of the well-known Young-Laplace formula for
the
force exerted by a tense interface.
The preceding two equations are the basis for all of our subsequent treatment
of
boundary motion.  With them, we can calculate ${\bf u}$ anywhere in the plane;
in
particular, the values of ${\bf u}$  on ${\cal C}$ give the interfacial
velocity.  Observe
that the boundary velocity is always determined {\it nonlocally}, and so
depends on the
shape of the entire domain.

Equation~(\ref{boundint}) for the fluid velocity can also be derived by a more
formal route~\cite{pozr.bk}:  In the spirit of textbook solutions of Laplace's
equation~\cite{jackson}, one begins by proving a reciprocal identity relating
two
arbitrary flows ${\bf u}$ and $\bf v$.
Choosing $\bf v$ to be proportional
to $G_{ij}$, one finds that the limiting value of ${\bf u}$ as it approaches
the
boundary $\cal C$ of an arbitrary region is
\begin{eqnarray}
u_j({\bf r})&=& -{1\over 2\pi\eta}
\int_{\cal C}\! ds \,  G_{ij}\left({\bf r}-{\bf r}(s)\right) f_i({\bf r}(s))
\nonumber \\
&&- {1\over 2 \pi} \int_{\cal C}\!ds \,  u_i({\bf r}(s))T_{ijk}
\left({\bf r}-{\bf r}(s)\right)n_k({\bf r}(s))~,
\label{boundint1}
\end{eqnarray}
where $f_i$ is the force that the fluid inside ${\cal C}$ exerts on ${\cal C}$,
$n_k$ is a component of the unit normal vector, and $T_{ijk} \equiv
-\delta_{ik}
P_j + \partial_k G_{ij} + \partial_i G_{kj}$ is the Green's function for the
stress tensor.  In the case in which the viscosities inside and outside ${\cal
C}$ are equal, one can readily combine the expressions for the limits from the
inside and the outside to recover equation~(\ref{boundint}).  When the two
viscosities are different, this approach does not generally work: $G_{ij}$ and
$T_{ijk}$ can depend on $\eta$, so terms containing the Green's functions for
the
inside and outside regions cannot necessarily be combined and cancelled.
Hence,
except in certain limits, our theory cannot immediately be extended to include
viscosity contrast.  Because of the difficulty of measuring the viscosities of
individual phases in the region of coexistence, the importance of such contrast
is usually not known.  In most systems of interest, dissipation in the
monolayer
is negligible compared to dissipation in the bulk~\cite{emann,kling-mcc},
suggesting that viscosity differences may not introduce too strong an effect.
Likewise, in the opposite limit in which the subfluid is completely ignored,
$G_{ij}$ does not depend on $\eta$.  The viscosity then enters the problem only
through the factors of $1/\eta$ in front of the integral in
equation~(\ref{boundint1}), and a little algebra makes it possible to deal with
variations in $\eta$.  It has been our experience that, in this case, the
correction term due to viscosity contrast usually vanishes to linear order.
Nonetheless, a more complete theory would allow for the possibility of
viscosity
differences between phases.

Before turning to the calculation
of $G_{ij}$, we lastly remark that our curve evolution dynamic produces a
gradient flow in configuration space.  That is, the energy
${\cal E}$ associated with an interface will always decrease monotonically in
time,
with $d{\cal E}/dt$ precisely given by the sum of the rates of viscous
dissipation
in the monolayer and in the subfluid.  To prove this, we begin by observing
that
${\bf \Delta f}= {\bf \Delta \sigma \cdot \hat {\bf n}} $, where ${\bf \Delta
\sigma}$
is the difference in stress tensors across the interface.  The time
derivative of the energy is then $d {\cal E}/dt = \int\! ds \, \left(\delta
{\cal
E}/ \delta {\bf r}\right)\cdot {\bf u} =
\int\! ds \hat{\bf n} \cdot {\bf \Delta \sigma} \cdot {\bf u}$.  The
divergence theorem allows one to transform this integral into an integral over
the plane; by using the dynamical equation~(\ref{modstokes}), one can then make
terms in ${\bf F}^S$ appear that can by similar arguments be written in terms
of
integrals over the subfluid volume.  Ultimately, one finds that
\begin{equation}
{d {\cal E}\over dt} = -2 \eta \int\! d^2{\bf r}\left(e_{ij}\right)^2 -
2 \eta' \int\! d^3{\bf r}' \left(e'_{ij}\right)^2 ~,
\end{equation}
where $e_{ij} = \left(\partial_i u_j + \partial_j u_i\right)/2$  is
proportional to the
viscous part of the stress tensor in the monolayer, and a similar definition
holds for $e'_{ij}$ in the subfluid.  Summation over repeated indices is
implied.  The two integrals give, respectively, the rates of viscous
dissipation in the monolayer and in the subfluid.  The result is hardly
surprising, but it has potentially important consqences for the pattern forming
properties of the model:  The system is constrained always to move ``downhill''
in the space of shapes, so it will tend to get caught in metastable minima, and
many shapes will be inaccessible to it.

\subsection{Calculation of the Green's Function}

To find $G_{ij}$, we begin by finding the velocity field in the
subfluid induced by an arbitrary flow in the monolayer; we will then proceed to
calculate ${\bf F}^S$ for this velocity field and finally to solve for
the Green's function.  The first task is greatly simplified by an
observation of Stone and McConnell~\cite{stone-mcc}, who showed that
when the monolayer velocity field is
incompressible the subfluid pressure $p'$ is constant.  Though their proof only
holds for an infinite subfluid, it can readily be extended to the case where
the
depth is finite; the result can also be verified independently starting from
the
expression for the subfluid velocity as an integral over the plane
$z=0$~\cite{pozr.bk}.  With a constant pressure, $\bbox{{\nabla}}p' = {\bf 0}$,
and each
component of  ${\bf u}'$ becomes harmonic.  Solving for ${\bf u}'$ is thus
reduced to an exercise in electrostatics.  Because the system is
invariant with respect to translations in the monolayer plane, the
drag force must take the form
\begin{equation}
{\bf F}^S({\bf r}_0) = {\eta'\over 4 \pi} \int\! d^2{\bf r} \,
K({\bf r}_0 -{\bf r}){\bf u}({\bf r})~, \label{convol-eqn}
\end{equation}
where $K$ is the derivative with respect to $z$ of the appropriate Green's
function for Laplace's equation.  Both $K$ and ${\bf u}$ are functions defined
on
the plane.  When the depth $d \rightarrow \infty$, $K$ is readily obtained by
the method of images; the extension to the case of a finite depth is
treated in Appendix~\ref{appA}.  It turns out to be convenient to
proceed in Fourier
space.  Denoting the Fourier transform (in two dimensions) of a function by a
hat and adopting the convention that $f({\bf r}) = \int\! d^2{\bf q}
\hat f ({\bf q}) \exp(i {\bf q} \cdot {\bf r})$, we find that
\begin{equation}
\hat K ({\bf q}) = -{q\over \pi} \coth(qd)~, \label{k-eqn}
\end{equation}
whence by the convolution theorem
\begin{equation}
\hat {\bf F}^S({\bf q}) = -\eta' q \coth(qd) \hat {\bf u}({\bf q})~.
\end{equation}
The obvious next step is to take the Fourier transform of the Green's function
equation~(\ref{forcedeqn}).  Making use of the fact that
$\bbox{\nabla}\cdot {\bf u}= 0$, one can obtain an expression for
$\hat{P_j}$ and thus show that
\begin{equation}
\hat G_{ij}({\bf q}) =  {q^2 \delta_{ij} - q_i q_j\over \pi q^2
\left[q^2 + a q \coth(qd) \right]}~. \label{gij-eqn}
\end{equation}
The parameter $a$ is
the ratio of the viscosities of the subphase and the monolayer,
\begin{equation}
a \equiv {\eta'\over \eta}~,
\end{equation}
and has dimensions of inverse length. This parameter plays a fundamental role
in all subsequent
analyses.  When multiplied by an appropriate
length scale, it will be the governing dimensionless parameter of the problem.

The form of the expression for $\hat G_{ij}$ suggests that we define the
differential operator
\begin{equation}
D_{ij} \equiv -{1\over \pi}\left(\delta_{ij}\nabla^2 - \partial_i
\partial_j\right)
\end{equation}
so that we can write $G_{ij}$ in terms of a single scalar function as
\begin{equation}
G_{ij} = D_{ij} H({\bf r})~,
\end{equation}
with
\begin{equation}
\hat H({\bf q}) = {1\over q^2 \left[q^2 + a q \coth(qd) \right]}~.
\end{equation}
In principle, we could now invert this transform and calculate $G_{ij}$.  In
practice, the result would be so cumbersome as to be useless.  Instead, in the
next section we will study the behavior of $\hat H$ in several different
limits.

\section{Limiting Cases}

In this section, we will consider the behavior of the model in the limits of
large and small $a$ and $d$.  Since both of these variables have dimensions, we
must compare them with some other quantity to have a meaningful notion of
``large" and ``small." The only candidate that presents itself in the present
formulation is the wavevector $q$.  If the system under consideration has a
single length scale $L$, then the most important contributions in Fourier space
are likely to come at $q \sim 1/L$, and the relevant dimensionless parameters
are
$d/L$ and $aL$.  This would be the case, for example, for a circular domain of
radius $R$ whose boundary was smooth on all smaller scales; we would then have
$L
\sim R$.  We caution, however, that $q \sim 1/L$ is at best a rough estimate
and
that $q$ must actually be allowed to range from zero to infinity.  Hence, great
care must be exercised in taking limits when there is more than one candidate
for
$L$.  This is particularly true when an approximation is valid only for large
enough $L$, for there is always the danger that the domain boundary will finger
or
develop roughness at smaller length scales.  For example, the
important length scale in the case of a circle subject to $n{\rm -fold}$
harmonic perturbations is not its radius $R$ but the wavelength $2 \pi R/n$;
whether or not a given approximation is valid thus depends on the mode one is
considering.

\subsection{Infinite Subfluid}

We begin by considering the case of a very deep subfluid, $qd
\rightarrow \infty$.  This is usually the experimentally relevant
limit, for typical troughs have depths on the order of millimeters, while the
monolayer
domains observed tend to be on the scale of tens, or at most hundreds of
microns~\cite{kling-mcc1}.  In this limit, $\coth(qd) \rightarrow 1$, and $\hat
H$
takes the simplified form
\begin{equation}
\hat H({\bf q}) = {1\over q^3 (q+a)}~.
\end{equation}
Note that in the limit of large $q/a$, $\hat H$  behaves as $q^{-4}$, while in
the limit of small $q/a$ it behaves as $q^{-3}$; these should determine the
behavior of $G_{ij}$ at small and large $r$, respectively.  If one lets $q/a
\rightarrow \infty$, or equivalently sets $a = 0$, one recovers the case of a
purely two-dimensional layer without any coupling to a bulk fluid.  Although
the
integrals required to take the inverse transform of $q^{-4}$ diverge in the
usual
sense, they can be dealt with by the theory of generalized functions
\cite{lighthill,jones}.  Essentially, all that is required is the introduction
of
a convergence factor like that commonly used to treat quantum-mechanical
scattering from a Coulomb potential.  One then finds that
\begin{equation}
H({\bf r}) = {\pi\over 2} r^2 \ln(r)~,
\end{equation}
and
\begin{equation}
G_{ij}({\bf r}) = -\delta_{ij}\ln(r) + {r_i r_j\over r^2}~,
\end{equation}
where $r_i$ is a component of ${\bf r}$.  This is precisely the two-dimensional
``Stokeslet" of fluid mechanics~\cite{pozr.bk,has-san}.  Although one should
generally not take the logarithm of a quantity with dimensions, in the above
equations this transgression turns out to be without consequences.  With $a=0$,
the system has no intrinsic length scale, and we can chose to divide $r$ by
whatever length we please; the only change will be in an unimportant additive
constant corresponding to a Galilean transformation.

It is tempting to try to treat the limit in which dissipation in the subfluid
dominates in the same manner just used for the case in which it is negligible.
Unfortunately, if one blithely takes the inverse transform of $1/(a q^3)$,
obtaining $-2 \pi r/a$, and proceeds to calculate a
Green's function, clearly unphysical results emerge. In the simplest terms, one
can
argue that this occurs because the limit of large $q/a$ will break down at
large
distances, a feature that is acceptable for finite size domains, while the
limit
of small $q/a$ becomes invalid precisely at the small distances that one must
always consider.  More specifically, the limit $a \rightarrow \infty$
corresponds to neglecting the term $\eta \nabla^2 {\bf u}$ in the dynamical
equation~(\ref{modstokes}).  One is then left with a lower order equation for
which one is
allowed to impose fewer boundary conditions, so the integral
formulation of equation~(\ref{boundint}) can no longer hold.
We will argue in Section~\ref{large-a-sect} that these difficulties can be
circumvented with the judicious use of cut-offs.

To proceed further within the present framework, however,
one must deal with the form for $\hat H$ valid for arbitrary $a$.
A fairly involved expression for the inverse transform is obtained in
Appendix~\ref{appB}; it can be expanded about $a r=0$ to give
\begin{eqnarray}
H({\bf r})&=& {\pi\over a^2} \Biggl[{4 C -3 +
4 \ln(ar/2)\over 8} \left(a r\right)^2 \nonumber \\
&& - {2\over 9} \left(ar\right)^3
+{1-C-\ln(a r/2)\over 32}\left(ar\right)^4+\cdots\Biggr],
\label{hseries}
\end{eqnarray}
where $C \simeq 0.577$ is Euler's constant.  Since $D_{ij} r^2 = 2\delta_{ij}$,
the term proportional to $r^2$ will add a constant velocity to $G_{ij}$.
Unlike in the case $a=0$, such a constant now has a physical meaning; the
presence of the subfluid destroys Galilean invariance.

\subsection{Thin subfluid}

We next turn to the case of a very thin subfluid layer, $qd \rightarrow 0$ or
$d/L \rightarrow 0$.  Although no experiments have yet been conducted in this
regime, it seems plausible that it might be experimentally accessible.
In numerical studies of monolayer flow in canals, for example, Stone has
observed
that the effects of finite depth become important when $d \sim L$~\cite{stone}.
For the largest experimentally accessible monolayer domains, a
trough with a depth $d \sim 100 \mu{\rm m}$
would then be required; this seems mechanically conceivable, although
reflection
from the bottom of the trough might make visualization with some microscopy
techniques difficult~\cite{emann}.  Beyond the fact that its theoretical
treatment is less involved, this limit has the potential advantage that it
would
give the experimenter, in the depth $d$, an additional parameter that could be
controlled with a fair degree of precision.  For example, Klinger and McConnell
have reported the ability to set $d$ to within $1 \mu{\rm
m}$~\cite{kling-mcc1}.

To lowest order in $qd$, $\hat H$ takes the form
\begin{equation}
\hat H = {1\over q^2 \left(q^2+\lambda^2\right)}~, \ \ \ \ \
 \lambda^2 \equiv {a\over d} = {\eta'\over \eta d}~.
\label{thingreen}
\end{equation}
The parameter $\lambda$ plays the same role as $a$ in the infinite subfluid
problem.  The inverse transform of $\hat H$ can be taken without great
difficulty
and a Green's function obtained.  It turns out to be more instructive, however,
to back up several steps and to consider the function $K$ introduced in the
previous section.  Expanding the Fourier transform of $K$, inverting, and
taking
the convolution with ${\bf u}$, we obtain
\begin{mathletters}
\begin{equation}
\hat K ({\bf q})=-{1\over \pi d}\left[1+{1\over
3}\left(qd\right)^2+\cdots\right]
\end{equation}
\begin{equation}
K({\bf r}) = -{4\pi\over d}\left[\delta({\bf r})+{1\over 3}d^2
\nabla^2\delta({\bf r})+
\cdots\right]
\end{equation}
\begin{equation}
{\bf F}^S = -{\eta'\over d}\left[{\bf u}+{1\over 3} d^2\nabla^2 {\bf u}+
\cdots\right]~.
\end{equation}
\end{mathletters}
The first term in the expansion of ${\bf F}^S$ has been derived by
Stone via more heuristic arguments~\cite{stone-mcc2,stone}; it is also
the expression for ${\bf F}^S$ that one obtains by treating the
subfluid in the lubrication approximation.  In qualitative terms, the
series tells us that as $d$ increases the force exerted by the
subfluid develops an increasingly nonlocal character.  For very small
$d$, the no-slip boundary conditions on the bottom of the trough
completely dominate behavior and prevent the effects of motion in the
monolayer from propagating through the subfluid. As $d$ increases,
however, different parts of the adsorbed layer become more and more
able to communicate with each other through the subfluid.  In the
opposite limit of infinite depth, we thus expect that ${\bf F}^S$ will
depend on the velocity field throughout the monolayer.

We confine ourselves for the moment to considering only the leading term in the
above expansions. It is of course also possible to look at higher order
approximations, but they lack the internal consistency of the lowest order
dynamics.  In particular, they will not always yield a law of motion that is a
gradient flow.  In the first approximation, a comparatively simple equation
holds:
\begin{equation}
\eta \nabla^2 {\bf u}- \nabla p - {\eta'\over d} {\bf u}= {\bf 0}~.
\label{thineqn}
\end{equation}
This equation has the same form as the Laplace transform with respect to time
of
the linearized Navier-Stokes equation $\rho \partial {\bf u}/\partial t = \eta
\nabla^2 {\bf u}- \nabla p$ and has been studied in this
context~\cite{pozr.bk}; it has
also been used to model flow in porous media~\cite{howell}.  If we ignore
dissipation in the monolayer entirely compared with dissipation in the
subfluid,
the equation reduces to Darcy's law ${\bf u}\propto -\nabla p$,
which describes quasi-two-dimensional flow in Hele-Shaw
cells~\cite{gold-sci,gold-tse}.  The Green's function for Eq.
(\ref{thineqn}) has previously been calculated~\cite{pozr.bk} and can readily
be
obtained by the taking the inverse transform of equation~(\ref{thingreen}).
One
finds that
\begin{equation}
H({\bf r}) = -{2 \pi\over \lambda^2}\left[\ln\left(\lambda r\right)
+K_0\left(\lambda r\right)\right]~,
\end{equation}
and
\begin{eqnarray}
G_{ij}({\bf r})&=&-2\delta_{ij} \left[{1\over\left(\lambda r\right)^2}
-K_0\left(\lambda r\right)-{K_1\left(\lambda r\right)\over \lambda r}\right]
\nonumber \\
&&
+2{r_i r_j\over r^2}\left[{2\over \left(\lambda r\right)^2}
-K_0\left(\lambda r\right)-{2K_1\left(\lambda r\right)\over \lambda r}\right]~,
\end{eqnarray}
where $K_i$ is a modified Bessel function.  Note that in the limit
$\lambda r \rightarrow 0$, $H$ approaches the expression
obtained in the complete absence of a subfluid.  Close enough to a singularity,
the
presence of the subfluid will always be negligible because the higher order
derivative
$\nabla^2{\bf u}$ will always dominate ${\bf F}^S$.

\section{Linear Stability}
\label{linstab-sect}

While the present formulation of the monolayer hydrodynamics permits the
investigation of boundary motion far from any regular geometric shape, where
a fully nonlinear analysis is necessary, we focus here on
the calculation of linearized growth rates about several simple geometries.  To
develop
physical understanding, we will begin with an interface in the shape of a
straight
line; we will then turn to the more experimentally relevant case of a circular
boundary.

\subsection{Stability about a Line}

Consider an unperturbed interface that rests on the line $y = 0 $ and subject
it to a sinusoidal perturbation ${\bf r}(x) = \hat y(k) {\rm e}^{ikx}{\bf
e}_y$,
where ${\bf e}_y$ is a unit vector in the $y$-direction; we use the wavevector
$k$
to distinguish a one-dimensional Fourier transform from a two-dimensional
transform with wavevector $q$.
This gives rise to a force ${\bf \Delta f} = \hat f(k) {\rm e}^{ikx} {\bf e}_y$
which in turn causes a velocity at the interface $\hat u(k) {\rm e}^{ikx}
{\bf e}_y$,  with $\hat u  = \partial \hat y /\partial t$.  For a boundary
without internal structure,  ${\bf \Delta f}$ must always be normally directed,
so there is no possibility that it will  have an ${\bf e}_x$ component.  The
relationship of $\hat f$ to $\hat y $ will depend  on ${\cal E}[{\bf r}]$,
and its particular form is not of immediate interest in the present discussion.
Quite generally, though, we expect that $\hat u  \propto \hat f \propto
\hat y $.  Since $\hat f$ is already first order small, once it has been
calculated we may consider that the boundary takes its unperturbed shape.  In
those cases in which an analytic expression for $G_{ij}$ is known, we may then
simply calculate the velocity component as
\begin{equation}
\hat u (k) = -{\hat f(k)\over 4 \pi \eta} \int_{-\infty}^{\infty}\! dx
G_{22}(x {\bf e}_x){\rm e}^{ikx}~.
\end{equation}
We find that in the limit of negligible subfluid dissipation ($a\to 0$),
\begin{equation}
\hat u (k) = -{\hat f(k)\over 4 \eta \vert k \vert}~,
\label{grow1}
\end{equation}
while in the case of a thin subfluid,
\begin{equation}
\hat u (k) = -{\hat f(k)\over 2 \eta\lambda^2}
\left[\vert k\vert -{k^2\over \sqrt{k^2+\lambda^2}}\right]~,
\label{grow2}
\end{equation}
Once again the effects of the
subfluid are unimportant at small enough length scales:  As
$\vert k \vert/\lambda \rightarrow \infty$, the expression approaches that
valid in the absence of a subfluid.  Similarly, as $\vert k \vert/\lambda
\rightarrow 0$, we
recover the form that has previously been calculated starting from Darcy's
law~\cite{gold-sci,gold-tse}.

When a compact direct space form for $G_{ij}$ is not
known, one may still find the growth rates by remaining in Fourier space.
The method is presented in Appendix~\ref{appC} for stability about a circle;
the
results of a similar calculation for a line are plotted in
figure~\ref{figure2} in terms of a reduced growth rate $\sigma(k)
\equiv \hat{u}(k) \eta' / \hat{f}(k)$.
At present, we simply state that, for an infinite subfluid in the limit
$\vert k \vert/a \ll 1$,
\begin{equation}
\hat u (k) = -{\hat f(k)\over \pi \eta'}~.
\label{grow3}
\end{equation}
This is the same $k$ dependence one would expect if dissipation occurred
only at the boundary of the domain, instead of in the bulk fluid
developments of Section~\ref{large-a-sect}.

The $k$ dependence of these growth rates can be understood on the basis of
relatively simple arguments.  Suppose that $\hat H({\bf q}) \sim
q^n$; $n$ is determined by the number of derivatives
of ${\bf u}$ and of $p$ in the dynamical equation.  Then, we expect that
$\hat G_{ij} \sim q^{n+2}$.  To find the linearized growth rate about a line,
one must first take the inverse transform of $\hat G_{ij}$ in two dimensions,
then, in a rough sense, take a one-dimensional Fourier transform of the
resulting function.  On purely dimensional grounds, this will introduce an
additional factor of $q$.  Indentifying $q$ with the wavevector $k$ of the
perturbation, we then expect that $\hat u (k) \sim k^{n+3}$.  This is
indeed the case:  When $a=0$, $\hat H \sim q^{-4}$ and $\hat u (k) \sim
k^{-1}$, and similarly for the other limits.

\subsection{Stability about a Circle}

As in the previous section, we begin by considering a slightly perturbed domain
parametrized as ${\bf r}(\theta) = R(1 + \epsilon_n {\rm e}^{in\theta}){\bf
e}_r$,
where ${\bf e}_r$ is a radially-directed unit vector.  We expect a force
$f_n {\rm e}^{in\theta} {\bf e}_r$ and a normal
velocity component $u_n {\rm e}^{in\theta} {\bf e}_r$; $u_n = R
d\epsilon_n/dt$.
The force must again be normally-directed for a structureless interface.
Incompressibility
requires that the fluid velocity have a tangential component at the interface,
unlike
in the case of a line; this component does not, however, affect the evolution
of
the boundary's shape, so we will ignore it. To lowest order, we may still
consider that
the force acts at the unperturbed circle: Although there can now exist a zeroth
order
force, it must be independent of $\theta$ and so will not cause any fluid
motion, even
when acting at the perturbed interface.

Whereas for stability about a line we used a variety of direct space forms for
$G_{ij}$, here we shall instead derive one
expression valid for arbitrary $a$ and $d$.  We begin by observing that, using
polar coordinates and integrating over the polar angle, we may formally write
\begin{equation}
H({\bf r}) = 2 \pi \int_0^{\infty}\! dq {J_0(qr)\over q^2
\left[q + a \coth(qd)\right]}~.
\label{inth}
\end{equation}
Starting from this representation of $H$, straightforward but lengthy
manipulations
(described in Appendix~\ref{appC}) lead to an expression for the growth rate
$u_n$:
\begin{equation}
u_n = -{n^2 R f_n \over \eta} \int_0^\infty\! dw {J_n^2(w)\over w^2
\left[w + a R \coth(wd/R)\right]}~.
\label{circlegrow}
\end{equation}
This is the central result of this section and the analog of the growth rates
given by equations~(\ref{grow1}),~(\ref{grow2}), and~(\ref{grow3}) for
perturbations about a line.  If we let the subfluid be infinite and set
$\coth(w
d/R) =1$, we recover the expression previously obtained by Stone and
McConnell~\cite{stone-mcc}.  When $a = 0$ or $a R \rightarrow \infty$, the
integral can be evaluated in closed form.  One finds that
\begin{equation}
u_n = -{R\over 4 \eta} {\vert n\vert\over n^2-1}f_n \ \ \ \ \ \  (\vert n\vert
\ge 2)~,
\label{layer}
\end{equation}
when dissipation occurs only in the monolayer, and
\begin{equation}
u_n = -{4\over \pi \eta'} {n^2\over 4n^2-1}f_n \ \ \ \ \ \ (\vert n \vert \ge
2)~,
\label{subflu}
\end{equation}
when the subfluid dominates. The leading
corrections to these expressions for finite $a R$ can also easily be computed.
In
both cases, the first correction tends to decrease the magnitude of the growth
rate.  This is not surprising:  To zeroth order, we entirely ignored
dissipation
in the subfluid, in the one case, and in the monolayer, in the other.  The next
term, by accounting for these additional sources of energy loss, increases the
total amount of damping and so slows down the dynamics.

One may verify that in the limit $n\to \infty$,
 $R\to \infty$, with $k=n/R$ fixed, the results (\ref{layer}) and
(\ref{subflu}) for a circle
tend to the growth rates about a straight line (\ref{grow1}) and (\ref{grow3}).
For $n$ large enough, the curvature of the circular boundary is unimportant on
the scales over which the forces and velocities vary, and the boundary acts
essentially like a line.  Hence, the length scale with which one must compare
$a$ is not the circle's radius $R$ but the wavelength of the perturbation,
which
is proportional to $1/k=R/n$.
The expression (\ref{layer}) is the appropriate approximation for small $a
R/n$,
while (\ref{subflu}) is more accurate for large $a R/n$.
This dependence on $n$ of the dominant source of dissipation can be seen in
figure~\ref{figure3}, where the exact reduced growth rate $ \sigma_n
\equiv u_n \eta' / f_n $ is compared with the two limiting forms. For fixed $a
R \gg 1$, the growth rate is roughly
independent of $n$ for small $n$ but decays like $1/n$ for large $n$.
The crossover occurs where the curves given by equations (\ref{layer})
and (\ref{subflu}) intersect, at $a R/n \sim 1$.  Except for a
multiplicative factor, the growth rate about a line is a function only
of $a/k$.  This is not precisely the case for a circle, but, even for
$n$ small, many important quantities depend essentially only on $a
R/n$.  For example, figure~\ref{figure4}\ plots versus $a R/n$ the
fractional difference between the exact growth rate given by equation
(\ref{circlegrow}) and the approximation of equation (\ref{subflu})
for a number of different values of $a R$ and $n$.  To a good
approximation, all of the points fall on the same
curve.  Actually, this collapse occurs over a much wider range of $aR/n$ than
shown in the figure, even when the fractional error is greater than ${\cal
O}(1)$.
Figure~\ref{figure4}\ also gives an idea of the error involved
in using the approximate expression for the growth rate.  We see that
for $a R$ as high as 100, the approximation is accurate to within 5\%
for the first few modes, with error increasing linearly as $n/a R$ for
larger $n$.  This asymptotic dependence of the error
may also be derived directly by expanding the integrand of
equation~\ref{circlegrow}~\cite{howard}.

\subsection{Example: Dipolar Forces}

In previous sections, we made no assumptions about the nature of the energy
functional ${\cal E}[{\bf r}]$ associated with the boundary.  Here we
undertake a sample calculation for the
functional of greatest experimental interest, that of a dipolar domain.  In
this case,
we associate with the boundary not only a line tension $\gamma$, but also the
electrostatic energy stored in the electric field created by the dipoles.  In
reality, this electric field exerts a force not at the boundary but on the bulk
dipolar fluid.
Under the assumption that the domain has a constant dipole density, however,
the electrostatic body force per unit volume
can be written as $-\nabla \phi$, where $\phi$ is
an appropriate potential energy.  After the introduction of a modified pressure
$p_{\rm m}=p+\phi$~\cite{pozr.bk}, the equations describing the bulk flow are
thus unchanged,
and the electrostatic interactions only enter through their effect on the
boundary conditions.  One can show that this effect is correctly incorporated
into our formalism
if one simply views the electrostatic energy of the dipolar domain as a
functional of its boundary's parametrization.
Several
equivalent forms exist for the energy ${\cal E}_d$ of an arbitrarily shaped
domain
with (constant) dipole density $\mu$ per unit area
\cite{gold-mono,gold-pra,gold-sci,gold-tse,keller,cebers-mono,mcc-thin,deutch}. 
Of these, the most useful to us takes the form of the energy of interaction of two
current loops:
\begin{eqnarray}
{\cal E}_d [{\bf r}] &=& - {\mu^2\over h} \int_{\cal C}\! ds_1 \int_{\cal C}\!
ds_2
\hat {\bf t}(s_1)\cdot \hat {\bf t}(s_2) \nonumber \\
&&\qquad\qquad\qquad \times \Phi\left(\vert {\bf r}(s_1)-{\bf
r}(s_2)\vert/h\right)~.
\end{eqnarray}
Here ${\cal C}$ is the curve parametrized by ${\bf r}$, $\hat{\bf t}$ is the
unit  tangent to ${\cal C}$, $h$ is the thickness of the monolayer, and
$\Phi(\xi) = \sinh^{-1}(1/\xi) + \xi - \sqrt{1+\xi^2}$.  Adding to the
electrostatic term the usual line tension energy $\gamma L$, where $L$ is the
length of the curve, we arrive at an expression for ${\cal E}[{\bf r}]$.  A
fair
amount of algebra then yields the force component $f_n$.  In the case of
monolayers, the thickness is of molecular size, $h \sim 10 {\rm \AA}$, and the
typical domain radius is $R \sim 10 \mu{\rm m}$, so we are justified in
taking the limit in which the aspect ratio $p \equiv 2 R/h \gg 1$.  It has then
been shown that~\cite{gold-mono}
\begin{eqnarray}
f_n &=&  {\gamma\over R} \Biggl[\Bigl\{1 - {1\over 2}N_B \ln\left({8R\over
eh}\right)
\Bigr\} \left(n^2-1\right) \nonumber \\
&&\qquad + {1\over 4} N_B \left(1-4n^2\right)
\sum_{j=2}^n{1\over 2j-1}\Biggr] ~.
\label{fn_dipoles}
\end{eqnarray}
Here the dipolar Bond number $N_B \equiv 2 \mu^2/\gamma$ gives the relative
importance of the electrostatic
and line tension forces.  Substituting this expression for $f_n$ into any of
the
growth rates calculated in the last section, one obtains a prediction for $u_n$
that can be compared directly with experiment.  With values of the
line tension on the order of $1\times 10^{-8}$ erg/cm~\cite{gold-mono},
a domain radius $R\sim 50 \mu m$, and subfluid viscosity $\eta'=1$ cp, we
obtain
interface velocities on the order of $1$ $\mu$m/sec for small $N_B$, and
considerably less near the branching instability.

\section{Dynamical Law for Motion Dominated by the Subfluid}
\label{large-a-sect}

Two important features of the limit in which $a L \rightarrow \infty$
have already been emphasized: First, the monolayer viscosity represents a
singular
perturbation that must always be taken into account near enough to a boundary
or to a
singularity.  Second, linear stability results suggest that this limit can be
partially understood in terms of an effective {\em local} dissipative force
that
opposes the boundary velocity at a given point on the interface.  It is the
purpose of
this section to use these two observations to find a simplified
boundary integral expression valid as $a L \rightarrow
\infty$.  The physical ideas that will motivate the
discussion are relatively straightforward:  The Green's function $G_{ij}$
deviates
appreciably from its asymptotic large $a$ form only when
 $ r \lesssim 1/a$.  For $a$ large enough,
this describes a very small region around the point where we wish to know the
boundary velocity, and it seems plausible that one might be able to neglect
the variation of physical quantities across this region.  Then, the
contribution
transmitted through the monolayer itself to the velocity at a point ${\bf r}$
will be
proportional to ${\bf \Delta f}({\bf r})$.  In this picture, the effective
force at the interface is thus a consequence of the extremely small length
scales over which dissipation in the monolayer is important.  These allow us to
take this dissipation to be essentially local compared with the dissipation in
the subfluid which retains its very nonlocal character.

To put these ideas into mathematical form, we begin by finding the limiting
forms of $G_{ij}$ for small and large $a r$.  These can be obtained by
straightforward differentiation of the corresponding limits of $H({\bf r})$,
derived in previous sections and in Appendix~\ref{appB}.  One finds that, for
$a r$
small, $G_{ij}$ behaves as
\begin{equation}
G_{ij}^S({\bf r}) = \left[{3\over 4} + C + \ln\left (a r/2\right) \right]
\delta_{ij} +
{r_i r_j\over r^2}~,
\end{equation}
while for $ar$ large the appropriate expression is
\begin{equation}
G_{ij}^L({\bf r})={2 r_i r_j\over a r^3}~.
\end{equation}
Note the important feature of $G_{ij}^L$ that, unlike most of the Green's
functions we
have examined, it does not contain a term proportional to
$\delta_{ij}$.   As a
first approximation, we will suppose that there is a sharp transition between
small and large $a r$ behavior.  That is, we will approximate the full Green's
function as
\begin{equation}
G_{ij}({\bf r}) = \left\{\begin{array}{ll}
G_{ij}^S({\bf r})~, & r < \upsilon/a \\
G_{ij}^L({\bf r})~, & r > \upsilon/a~,
\end{array}
\right.
\end{equation}
where $\upsilon$ is a constant of order unity that will be determined later.
The
expression for the velocity at a point ${\bf r}$ then becomes
\begin{eqnarray}
u_j({\bf r})&=&-{1\over 4 \pi \eta} \int_{\vert {\bf r} - {\bf r}(s)\vert >
\upsilon/a} \!\!\!\!\!\!\! ds \, G_{ij}^L({\bf r}-{\bf r}(s)) \Delta f_i({\bf
r}(s))
\nonumber \\ &&
-{1\over 4 \pi \eta} \int_{\vert{\bf r}-{\bf r}(s)\vert <
\upsilon/a} \!\!\!\!\!\!\! ds \, G_{ij}^S({\bf r}-{\bf r}(s)) \Delta f_i({\bf
r}(s)).
\end{eqnarray}

This expression can be written in an alternative form in the usual case in
which there are no tangential forces and one is only interested in the normal
velocity.  Then, the product $G_{ij}^L\Delta f_i$ is not
singular as ${\bf r}(s)$ approaches ${\bf r}$ because the only nonvanishing
term
in the sum  contains a factor of $[({\bf r}(s) - {\bf r}) \cdot \hat {\bf
n}]^2$
in the numerator.   Choosing the coordinate $s=0$ when ${\bf r}(s)={\bf r}$ and
expanding for small $s$, one finds that ${\bf r}(s)-{\bf r}\simeq \hat{\bf t}-
(1/2)\kappa \hat {\bf n} s^2 + \cdots \,$.
Hence,  $[({\bf r}(s) - {\bf r}) \cdot \hat {\bf n}]^2\propto s^4$, cancelling
the
singularity in the denominator of $G_{ij}^L$ and making it possible to extend
the
integral of $G_{ij}^L$ to zero.  For notational convenience, we pick axes such
that
$\hat {\bf n}$ and $\hat {\bf t}$ correspond to
${\bf e}_x$ and ${\bf e}_y$, respectively, when $s=0$.
Then, the normal component of ${\bf u}$ at ${\bf r}$ is just $u_1$, and we
can write
\begin{eqnarray}
u_1({\bf r})&=&-{1\over 4 \pi \eta} \int_{\cal C}\! ds G_{i1}^L({\bf r}-{\bf
r}(s))  \Delta f_i({\bf r}(s))  \nonumber \\
&&-{1\over 4 \pi \eta} \int_{\vert {\bf r}-{\bf r}(s)\vert < \upsilon/a} \!\!\!
ds
\Bigl[G_{i1}^S({\bf r}-{\bf r}(s))\nonumber \\
&&\qquad\qquad\qquad -G_{i1}^L({\bf r}-{\bf r}(s))\Bigr]
\Delta f_i({\bf r}(s))~.
\end{eqnarray}

Because $G_{ij}^L$ contains a factor of $1/a$, the first integral in the above
expression
will be proportional to $1/a$.  Thus, to leading order in large $a$, only
terms of order $1/a$ in the second integral are of interest.  At this level of
approximation, only constant and logarithmic terms in the integrand need be
retained---that is, forces and velocities are taken to vary very slowly across
the
region of integration.  Then, the second integral becomes
\begin{eqnarray}
-{1\over 4 \pi \eta}&&\int_{-\upsilon/a}^{\upsilon/a} \! ds \Delta f_1({\bf r})
\!
\left[\ln\left(a \vert s\vert /2 \right)+C+{3\over 4}\right]\nonumber \\
&=&{\Delta f_1({\bf r}) \upsilon\over 2\pi\eta'}
\left[\ln\left(\upsilon/2\right)+C-{1\over 4}\right]~,
\label{local-force}
\end{eqnarray}
where it should be emphasized that $\Delta f_1({\bf r}) =
{\bf \Delta f}({\bf r}) \cdot {\bf e}_x$ is the force component at the point
where we wish to calculate the velocity and so does not depend on $s$.
Hence, to leading order in large $a$, the second integral yields a local force.
We must still, however, specify
the numerical value of the proportionality constant $\upsilon [\ln(\upsilon/2)+
C-1/4]$ in equation~(\ref{local-force}), or, equivalently, of $\upsilon$
itself.
This can be done by demanding that the approximations of this section lead
to the same linear stability results as the limits as $a \rightarrow \infty$ of
the expressions valid for arbitrary $a$ already obtained.  Thus we ask that
$\upsilon$ be such that $\hat{u}(k) = -\hat{f}(k)/\pi \eta'$ (Eqn.~\ref{grow3})
for stability
about a line, and similarly for other geometries.
This requirement determines the unique value
$\upsilon \approx 2.88$.
Likewise, with this
choice of $\upsilon$, one can show that the proposed dynamical law preserves
the
area of domains and that the interfacial velocity of an arbitrarily shaped
domain will vanish when the force is constant and normally-directed, both very
desirable features.  Substituting for $\upsilon$, we find the approximation to
equation~(\ref{boundint}) that is the central result of this section:
\begin{equation}
u_1({\bf r}) = -{\Delta f_1({\bf r}) \over \pi \eta'}-{1\over 4 \pi \eta}
\int_{\cal C}
\! ds ~G_{i1}^L({\bf r}-{\bf r}(s)) \Delta f_i({\bf r}(s))~.
\label{large-a-int}
\end{equation}
The first term can be associated with dissipation in the monolayer and is
entirely local, while the second corresponds to dissipation in the
subfluid.
Note that the factor of $1/a$ in $G_{ij}^L$, when multiplied by the $1/\eta$
in front of the integral, yields $1/\eta'$, so the velocity does not
depend at all on the monolayer viscosity.

It remains to clarify when the results of this section are expected to
hold---that is, what it means for $a$ to be large.
By considering the corrections of order $1/a^2$ to the
expression~(\ref{local-force}),
 one can convince oneself that these are negligible only when
\begin{equation}
a \gg \kappa \ \ \ \ {\rm and} \ \ \ \  a \gg {\Delta f'_1({\bf r})
 \over \Delta f_1({\bf r}) }~,
\end{equation}
where $\kappa$ is the curvature,
$\Delta f'_1 \equiv d({\bf \Delta f}({\bf r(s)})
\cdot {\bf e}_x)/ds$, and
both quantities are evaluated at ${\bf r}$.  These expressions are not
surprising:  For smooth interfaces, the curvature will be of the order of
$1/L$,
where $L$ is the size of the domain; $\Delta f'_1/\Delta f_1$ gives the length
scale
over which the force varies.  As we have already pointed out, both are
important length scales in the system.

Unlike the Green's function valid for arbitrary $a$, $G_{ij}^L$ has a
compact analytic expression in direct space.  The
formula~(\ref{large-a-int}) thus presents considerable advantages for
analytic and especially for numerical calculations.  Such a simplification is
all the more welcome because almost all experiments to date have been
performed on systems where the subfluid viscocity dominates.
Admittedly, these improvements come at the expense of
an approximation that is not perfectly controlled, but that nonetheless appears
physically reasonable.

\section{Evolution of the Perimeter}
\label{perimeter-sect}

Finally, we provide a simple illustration of the use of the present formulation
for
calculations in the nonlinear regime.  We consider a domain with only line
tension energy, ${\cal E}=\gamma L$, and calculate the time derivative of its
perimeter $L$.  This quantity is of particular interest because $L$
can be observed directly in experiments.

Parametrizing the curve ${\cal C}$ in the standard way as ${\bf
r}(\alpha)$, and recalling that $\delta L/\delta {\bf r}(\alpha) = - d\hat {\bf
t}/d\alpha$, we can write
\begin{eqnarray}
{d L\over d t}&=&-\gamma \int_{\cal C}\! d\alpha {d \hat {\bf t}\over d\alpha}
\cdot
{\bf u}(\alpha)\nonumber \\
&=&-{\gamma \over 4\pi\eta} \int_{\cal C}\! ds_1 \int_{\cal C}\! ds_2
\kappa(s_1)
n_i(s_1) \nonumber \\
&&\qquad\qquad \times G_{ij}({\bf r}_{12}) \kappa(s_2) n_j(s_2)~,
\end{eqnarray}
where ${\bf r}_{12}= {\bf r}(s_1)-{\bf r}(s_2)$, $\kappa$ is the curvature,
and $s$ is the arclength parameter.  Since ${\cal E} \propto L$, it will always
be true that $dL/dt < 0$; the proof is identical to that already given for an
arbitrary energy functional. The Green's function $G_{ij}$ can be any of the
several forms we have already presented.
One could equally use the approximate form~(\ref{large-a-int}) of the previous
section for the velocities, finding that
\begin{eqnarray}
{dL\over dt}&=&-{\gamma \over \pi \eta'} \int_{\cal C} \! ds \kappa^2 \nonumber
\\
&&-{\gamma \over 4\pi\eta} \int_{\cal C}\! ds_1 \int_{\cal C}\! ds_2
\kappa(s_1)
n_i(s_1)\nonumber \\
&&\qquad\qquad \times G_{ij}^L({\bf r}_{12})\kappa(s_2) n_j(s_2)~,
\end{eqnarray}
when the monolayer viscosity is small.  These expressions for $d L/dt$ are
the analogs of the area-conserving ``curve-shortening equation'' previously
derived under the assumption that dissipation occurs only at the domain
boundary~\cite{gold-pra}.  This simpler equation has already been fruitfully
compared with experiment~\cite{seul-dyn,gold-mono}, and we hope that our
extensions will
likewise prove useful in analyzing data.

\section{Discussion}
\label{discussion-sect}

Equation~(\ref{boundint}) encapsulates the focus of this paper:  It
provides a general prescription for finding the velocity of the boundary of an
arbitrarily shaped monolayer domain.
We have discussed the form this equation
takes in several limiting cases and have shown that it can be used to make
quantitative predictions that can be compared with experiment.
In particular, section~\ref{large-a-sect}
suggests an approach to the limit in which dissipation in the monolayer itself
is
small; since most experiments have been conducted in this
regime~\cite{emann,kling-mcc}, its dynamics are of particular interest.  Our
approximation to equation~(\ref{boundint}) has the interesting feature
that it contains a local term whose effect is equivalent to that of a drag
force
acting directly on the interface.  From a theoretical perspective, it is clear
that our somewhat {\it ad hoc} treatment of this limit leaves a number of open
questions.  For example, the issue of whether the dynamical
law~(\ref{large-a-int}) gives precisely a gradient flow demands further
elucidation, as does the application of our arguments to the case of a thin
subfluid, which seems to reduce to Darcy's law when the subfluid viscosity
dominates.  In general, a more mathematically rigorous treatment is desirable.
These problems are currently under active study~\cite{dkl}. The logical next
step would be to undertake a detailed numerical investigation of relaxation and
pattern formation in the strongly nonlinear regime; techniques that have been
developed for similar problems~\cite{pozr.bk} should make such studies
possible.  To provide a complete test of the theory, it likewise seems useful
to
conduct experiments on systems in which the monolayer viscosity is sufficiently
large that it can be measured independently~\cite{jarvis,sacch}.  Experiments
carried out on a shallow trough might also be of interest.  Finally, there are
several physical effects that our theory does not pretend to include.  Foremost
among these are, first, the possibility that one of the monolayer phases might
have a
significant compressibility, and, second, the presence of thermal fluctuations
in domain
shape.  The inclusion of thermal fluctuations, in particular, might have a
significant qualitative effect on the pattern-forming properties of the model.
We also have made no attempt to treat the dynamics of the various liquid
condensed phases that show hexatic or other long-range order.  Clearly, such
order greatly increases the complexity of the problem.

\section{Acknowledgments}

We thank M. Gannon, E. K. Mann, and M.J. Shelley for helpful
discussions.  We are especially grateful to H. A. Stone for a careful
reading of a late version of the manuscript and for making available
the numerical routines used in generating figure 4.  This work was
supported in part by a Barry M. Goldwater Scholarship and an NSF
Graduate Research Fellowship (DKL), and by an NSF Presidential Faculty
Fellowship DMR93-50227 and the Alfred P. Sloan Foundation (REG).

\appendix

\section{Details of the Derivation of the Green's Function}
\label{appA}

This appendix gives a few of the steps that were omitted in the main
text in the derivations of
equations~\ref{convol-eqn} and~\ref{k-eqn}.  Since the components of
$\bf u'$ are harmonic, they can be
represented in the usual manner as
\begin{equation}
u'_i({\bf r}_0) = - \frac{1}{4 \pi} \int d^2{\bf r} \, u_i({\bf r})
\frac{\partial {\cal G}({\bf r},{\bf r}_0) }{\partial z} \; ,
\label{electro-green}
\end{equation}
where $\cal G$ is a Green's function for Laplace's equation that
vanishes on the plane $z = 0$, and the integral is taken over this
plane~\cite{jackson}.  The boundary conditions are that ${\bf u}'_{\perp} =
{\bf u}$
at $z = 0$, so the two dimensional velocity components $u_i$ appear in
the integrand.  The Green's function $\cal G$ is obtained by an
eigenfunction expansion in Cartesian coordinates.
With the hat denoting a Fourier transform with respect to
the variables $x-x_0$ and $y-y_0$, one finds that
\begin{equation}
\hat{\cal G}({\bf q},z,z_0) = -\frac{1}{\pi q} \frac{\sinh (q z_>) \sinh [q
(z_< + d)]
}{\sinh (qd) } \; ,
\end{equation}
where $z_>$ and $z_<$ are respectively the larger and smaller of $z$
and $z_0$.  Since ${\bf F}^{\rm S} = - \eta' \partial {\bf u'}_\perp /
\partial z$, comparison of equations~\ref{convol-eqn}
and~\ref{electro-green}  indicates that
$\hat{K} = \partial^2 \hat{\cal G}/ \partial z \partial z_0$, with the
derivatives evaluated at $z = z_0 = 0$.  The expression~\ref{k-eqn}
for $\hat{K}$ follows immediately.

\section{Inversion of $\hat H$}
\label{appB}

In this appendix, we obtain an analytic expression for the inverse transform of
$\hat H = 1/[q^3 (q+a)]$.  Begin by expanding in partial fractions:
\begin{equation}
\hat H({\bf q}) = {1\over a q^3}-{1\over a^2 q^2}+{1\over a^3 q}-{1\over a^3
(q+a)}~.
\end{equation}
The inverse transforms of the first three terms are known~\cite{jones}.
To invert the fourth, we use the general result that
${\cal F}^{-1}[g({\bf q})] = -(1/r^2) {\cal F}^{-1}[ \nabla^2_q g({\bf q})]$,
where
${\cal F}^{-1}$ denotes an inverse Fourier transform and the derivatives in the
Laplacian are taken with respect to $q$.  We can then write
\begin{eqnarray}
H({\bf r})&=&{2 \pi\over a^2} \Bigl[-\xi+\ln\xi+{1\over \xi}\nonumber \\
&&\qquad +{1\over \xi^2} \int_0^\infty \! dx J_0(\xi x)
{x-1\over (x+1)^3}\Bigr]~.
\label{hr}
\end{eqnarray}
Here, we have expressed the result in a dimensionless form by setting $\xi =
ar$
and have performed the integral over the polar angle in the inverse transform
of
$\nabla^2_q [1/(q+a)]$.  The integral in the fourth term converges in the usual
sense
and so can be evaluated numerically.  It can also be expressed analytically in
terms of a rather lengthy expression involving generalized hypergeometric
functions $_pF_s$.   These have known Taylor expansions that converge
everywhere in the complex plane~\cite{g-r}.
The expansions lead directly to the series~(\ref{hseries}).  It can also
be shown that to leading order for large $\xi$, $ H({\bf r}) = -2 \pi \xi/a^2$,
as one would expect from limiting form of the Fourier transform for small $q$.

\section{Linear Stability Calculation for a Circle}
\label{appC}

We take as our starting point the integral expression~(\ref{inth}) for $H({\bf
r})$.
Differentiating under the integral sign, we find that
\begin{eqnarray}
G_{ij}({\bf r})&=& 2 \pi \int_0^{\infty}\!dq {D_{ij}J_0(qr)\over
q^2\left[q+a\coth(qd)\right]}
\nonumber \\
&=&-2 \int_0^\infty \!{dq\over q^2 [q+ a \coth(qd)]}
\Bigl[{r_i r_j\over r^3} q J_0'(qr)\nonumber \\
&&\qquad\qquad +\left(\delta_{ij}-{r_i r_j\over r^2}\right)q^2 J_0''(qr)\Bigr]
{}.
\label{gijint}
\end{eqnarray}
In order to find $u_n$, we calculate
the velocity at $\theta = 0$.  Parametrizing the curve as ${\bf r}(\theta) = R
(\cos \theta {\bf e}_x+ \sin \theta {\bf e}_y)$, we may write the velocity in
the
usual manner as the integral around the curve of the Green's function
multiplied
by the force:
\begin{eqnarray}
u_n &=& -{Rf_n\over 4\pi\eta} \int_0^{2 \pi}\!\!\! d\theta
G_{1j}\left(R\left[\left(1-\cos\theta\right)
{\bf e}_x - \sin \theta {\bf e}_y\right]\right)\nonumber \\
&&\qquad\qquad \times {\rm e}^{in\theta} {\bf e}_{r,j}~,
\end{eqnarray}
where ${\bf e}_{r,j}$ is a component of ${\bf e}_r$.  Substituting the integral
expression~(\ref{gijint}) for $G_{ij}$, setting $w \equiv q R$, and performing
some
algebraic manipulations, one finds that
\begin{eqnarray}
u_n&=&{f_n\over \pi \eta R} \int_0^{\infty}\!\!{dq\over q^2 \left[q+a
\coth(qd)\right]}
\int_0^\pi\!\! d\phi \cos(2 n \phi) \nonumber \\
&&~\times\left[{w J_1(2w\sin\phi)\over 2\sin\phi}-w^2 \cos^2\phi
J_0(2w\sin\phi)\right]~.
\end{eqnarray}
The identity $\pi J_n(z)^2 = \int_0^\pi d\phi J_0(2 z \sin \phi) \cos(2
n \phi)$~\cite{g-r} and two integrations by parts enable one to evaluate the
integral with respect to $\phi$.  The growth rate given in
section~\ref{linstab-sect} is then obtained by changing variables of
integration
from $q$ to $w$.  The same
approach can also be applied to other geometries.

\begin{figure}
\caption{Schematic of the system studied, in which an interface ${\cal C}$
exists within a monolayer of viscosity $\eta$ resting on top of a subphase
with viscosity $\eta'$ and depth $d$.
\label{figure1}}
\end{figure}

\begin{figure}
\caption{Comparison of the reduced growth rates $\sigma(k) \equiv\break
\hat u(k) \eta'/\hat f(k)$ about a line in the limits of a very thin subfluid
and of an infinite subfluid. Both curves are for a system with $a = 10
\mu{\rm m}^{-1}$.  The dotted line was calculated using equation
(\protect{\ref{grow2}}) with $d = 10 \mu{\rm m}$; the solid curve was obtained
by the
methods of Appendix B in the limit $d \rightarrow \infty$.  The two curves
approach each other and drop off like $1/k$ as $k \rightarrow \infty$,
 but have markedly different behavior
for values of $1/k$ on the order of typical domain length scales.
\label{figure2}}
\end{figure}

\begin{figure}
\caption{Reduced growth rate $\sigma_n \equiv u_n \eta'/f_n$
versus the mode number $n$ for a circular domain with $a R=25$ resting
on an infinitely deep subfluid.  The solid line
gives the exact value, calculated from equation~\protect\ref{circlegrow}.  The
downward sloping dotted line gives the expression valid when dissipation in the
subfluid is negligible (equation~\protect\ref{layer}), while the other dotted
line
plots the expression valid in the opposite limit
(equation~\protect\ref{subflu}).
The exact growth rate is always less than either of the approximations, but
approaches the limiting expressions as $n \rightarrow 2$ and $n \rightarrow
\infty$.  \label{figure3} }
\end{figure}

\begin{figure}
\caption{Fractional error $[\sigma_n(a\to\infty) - \sigma_n]/\sigma_n $
versus $a R/n$ for a
circular domain resting on an infinite subfluid.  Here $\sigma_n
\equiv u_n \eta' / f_n$ is the
reduced growth rate calculated from the exact
expression~(\protect\ref{circlegrow}),
while $\sigma_n(a\to\infty)$ is obtained from the approximation
(\protect\ref{subflu}) valid when dissipation in the monolayer can be
neglected.
The solid line gives the best-fit power law,
$[\sigma_n(a\to\infty)-\sigma_n]/\sigma_n \simeq 1.4 /(a
R/n)$. \label{figure4}}
\end{figure}


\begin{references}
\bibitem{end-notea} It is well-known that high Reynolds number flows can be
treated
as incompressible when typical flow velocities are much less than the speed of
sound in the fluid.  One can readily show that the analogous criterion for low
Reynolds number flow in which dissipation in the subfluid dominates is $\eta' U
\beta \ll 1$, where $U$ is a typical velocity and $\beta$ is the
compressibility
of the monolayer.  Denser liquid-like phases easily satisfy this condition,
in agreement with the observation that domain areas generally remain
constant to within experimental error~\cite{seul-dyn,emann}.
It is not usually possible, however, to measure the compressibility of
the considerably less dense gas-like
phases that are present in some experimental systems, so one can legitimately
question whether it is valid to treat them as incompressible.
\end{references}
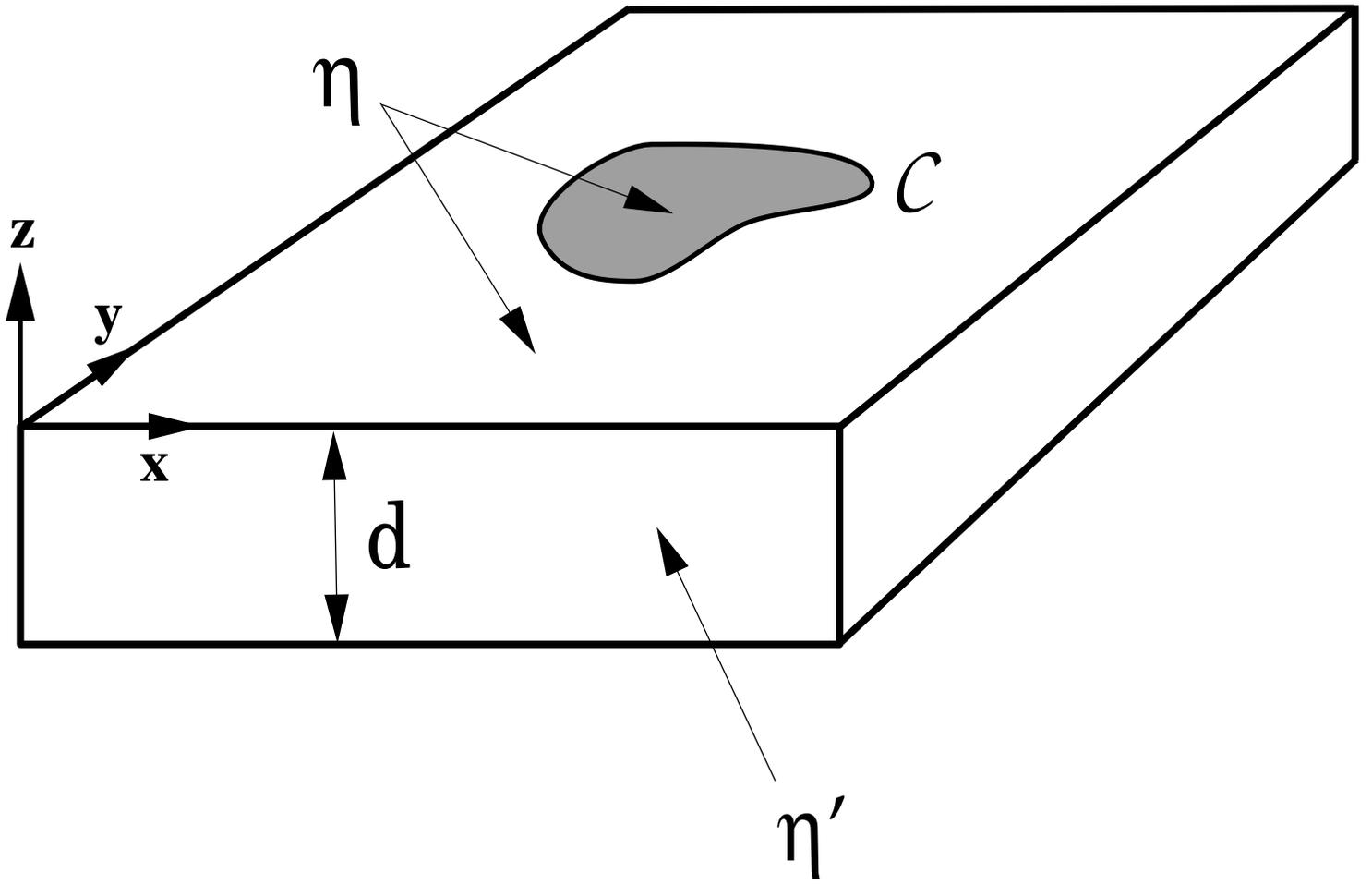
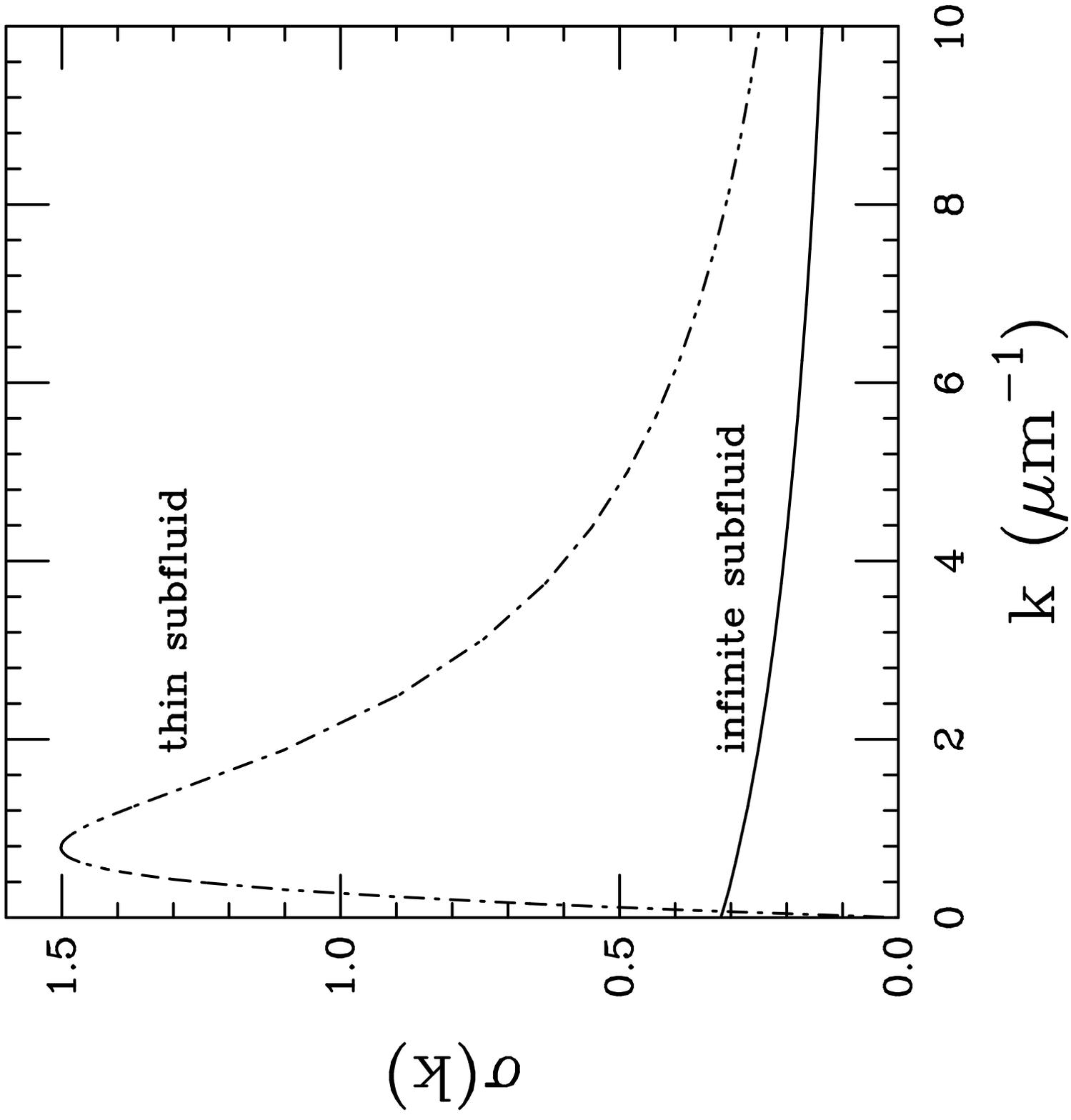
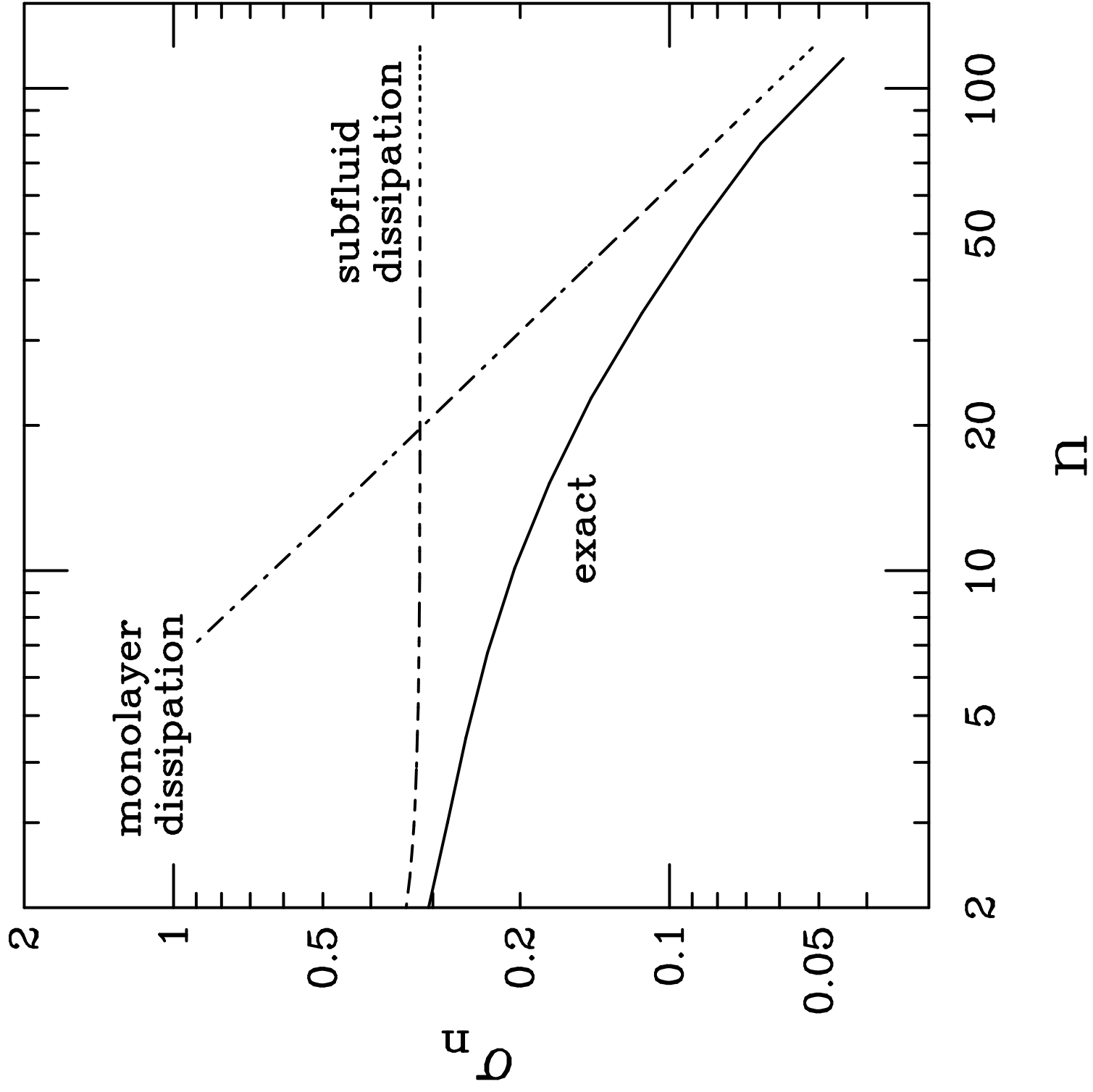
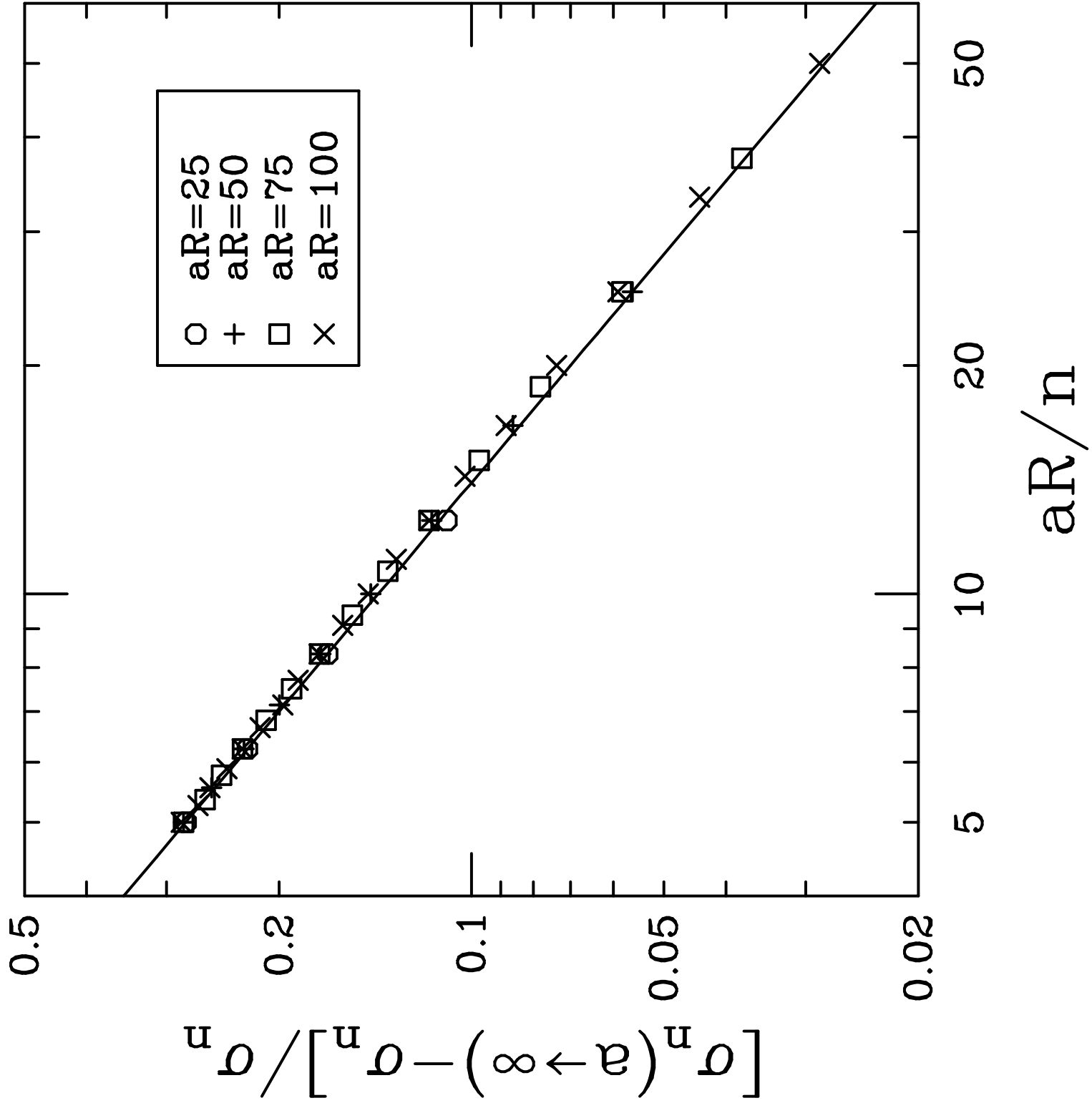
\end{document}